\documentstyle[12pt]{article}\textheight 230mm\textwidth 150mm
            \newcommand{\be}{\begin{eqnarray}}
            \newcommand{\ee}{\end{eqnarray}}
           \newcommand{\eel}[1]{\label{#1}\end{eqnarray}}
\newcommand{\e}[1]{\label{e:#1}\end{eqnarray}}
     \newcommand{\eg}{{\em e.g.\ }}
            \newcommand{\ie}{{\em i.e.\ }}
            \newcommand{\ga}{{\gamma}}
 
            \newcommand{\la}{{\lambda}}
  \newcommand{\La}{{\Lambda}}
            
            \newcommand{\del}{{\delta}}
 \newcommand{\om}{{\omega}}

\newcommand{\cL}{{\cal L}}

            \newcommand{\beq}{\begin{quote}}
            \newcommand{\eq}{\end{quote}}
            
            \newcommand{\al}{\alpha}
            \newcommand{\ben}{\begin{enumerate}}
            \newcommand{\een}{\end{enumerate}}
            \newcommand{\bit}{\begin{itemize}}
            \newcommand{\ei}{\end{itemize}}
    	\newcommand{\nn}{\nonumber}
            \newcommand{\r}[1]{(\ref{e:#1})}
            \newcommand{\edfl}[1]{\Label{#1}\end{df}}

\newcommand{\cD}{{\cal D}}

\newcommand{\ve}{{\varepsilon}}

\newcommand{\dif}{{\partial}}
\newcommand{\half}{\frac{1}{2}}
	
	\newcommand{\ldif}{{\stackrel{\leftarrow}{\partial}}}
\newcommand{\ldel}{{\stackrel{\leftarrow}{\delta}}}

\begin{document}
\begin{titlepage}

\vspace*{40mm}
\begin{center}{\LARGE\bf
Superfield algorithms for}\end{center}
\begin{center}{\LARGE\bf topological field theories}\end{center}
\vspace*{3 mm}
\begin{center}
\vspace*{3 mm}

\begin{center}Igor Batalin\footnote{On leave of absence from
P.N.Lebedev Physical Institute, 117924  Moscow, Russia\\E-mail:
batalin@td.lpi.ac.ru.} and Robert
Marnelius\footnote{E-mail: tferm@fy.chalmers.se}
 \\ \vspace*{7 mm} {\sl Department of Theoretical Physics\\ Chalmers University of Technology\\
G\"{o}teborg University\\
S-412 96  G\"{o}teborg, Sweden}\end{center}
\vspace*{25 mm}
\begin{abstract}
A superfield algorithm for master actions of a class of gauge field theories
including  topological ones in arbitrary dimensions is presented
generalizing a previous treatment in two dimensions.  General forms for master
actions in superspace  are given, and  possible  theories are  determined
by means of
a ghost number prescription and the master equations. The resulting master
actions
determine the original actions together with their gauge invariances.
Generalized
Poisson sigma models in arbitrary dimensions are constructed by means of this
algorithm, and other simple  applications in low dimensions  are  given  including a derivation of the non-abelian
Chern-Simon model.
\end{abstract}\end{center}\end{titlepage}
\section{The basic formulation}
In a previous paper \cite{Sigma} we gave
 a superfield algorithm for a class of master actions in two dimensions  by
means
of which we derived generalized Poisson sigma models. (A superfield form of
the master action for the ordinary Poisson sigma models \cite{Ikeda,SS} was
given
by Cattaneo and Felder \cite{CF}.)  In this communication we generalize this
algorithm  to arbitrary dimensions. This provides  then  for a general
framework for
 topological field
theories and generalizations. These models all share the following
properties: i)  The
equations of motion are of first order in the derivatives. ii) They are directly
defined
 in terms of their corresponding master actions. iii) There are general
simple rules for how the original actions are obtained from these master
actions.
iv) The  quantum theory of the models  are obtained by a gauge fixing of
the master
actions. For other related works see
also refs.\cite{AKSZ,GD,CR,Park,CF1,CF2,BBD}.

 The master actions corresponding to an $n$-dimensional
 field theory will be entirely expressed in terms of fields on a
supermanifold of dimension $(n, n)$, \ie half of the coordinates are
bosonic, even
ones, and half are fermionic, odd ones. To begin with we consider a class of
master actions expressed in terms of pairs of superfields. They are
\be
&&\Sigma[\Phi, \Phi^*]=\int d^{n}u d^{n}\tau \cL_n(u,\tau),
\e{1}
where the Lagrangian densities $\cL_n(u,\tau)$ are given by
\be
&\cL_n(u,\tau)&=\Phi^*_A(u,\tau)D\Phi^A(u,\tau)(-1)^{\ve_A+n}-
S(\Phi(u,\tau),\Phi^*(u,\tau)),
\e{2}
where $u^a$, $a=1,2,\ldots,n$ are bosonic coordinates on the base space and
$\tau^a$,
$a=1,2,\ldots,n$, the corresponding fermionic ones.    $D$ is
the odd de Rham differential
\be
&&D\equiv \tau^a\dif_a, \quad \dif_a\equiv{\dif\over\dif u^a}\quad \Rightarrow
\quad D^2=0.
\e{3}
Since $\Sigma$ is required to be even and since the measure $d^n\tau$  is
even or odd
depending on whether $n$ is even or odd, it follows that the Grassmann
parity of the
Lagrangian density is $n$, \ie $\ve(\cL_n(u,\tau))=n$. This implies that
$\ve(S)=n$
and
$\ve(\Phi^*_A)=\ve_A+1+n$, where
$\ve_A\equiv\ve(\Phi^A)$.  (For even
$n$ the form of
$\cL_n$ in
\r{2} is a generalization of the two dimensional treatment in
\cite{Sigma}, and for odd $n$ it is a generalization of the
one-dimensional  expression given in
\cite{GD,BBD}. Such superfield forms for master actions for topological field
theories were also proposed in
\cite{AKSZ} (see \eg \cite{GD}).)

The master actions
\r{1}  are required to satisfy the classical master equation
\be
&&(\Sigma,\Sigma)=0,
\e{4}
where the antibracket is defined by
\be
&(F, G)\equiv& \int
F{\ldel\over\del\Phi^A(u,\tau)}(-1)^{n\ve_A}d^{n}u
d^{n}\tau{\stackrel{\rightarrow}{\del}\over\del\Phi^*_A(u,\tau)}G-\nn\\&&-(F
\leftrightarrow
G)(-1)^{(\ve(F)+1)(\ve(G)+1)},
\e{5}
which in turn may be defined by \cite{EW}
\be
&&(F,G)=\Delta(FG)(-1)^{\ve_F}-(\Delta F)G(-1)^{\ve_F}-F(\Delta G)
\e{102}
in terms of the basic  odd, nilpotent $\Delta$-operator
\be
&&\Delta\equiv\int d^{n}u d^{n}\tau
(-1)^{(n+1)\ve_A}{\del\over\del\Phi^A(u,\tau)}{\del\over\del\Phi^*_A(u,\tau)}.
\e{6}
This $\Delta$-operator also differentiates the antibracket \r{5} according to
Leibniz' rule, \ie
\be
&&\Delta(F,G)=(\Delta F, G)+(F, \Delta G)(-1)^{\ve_F+1}.
\e{601}
 The functional derivatives in
\r{5} and
\r{6} are defined by the properties
\be
&&{\del\over\del\Phi^B(u,\tau)}\Phi^A(u',\tau')=
\del^A_B\del^n(u-u')\del^n(\tau-\tau')=\Phi^A(u,\tau)
{\stackrel{\leftarrow}{\del}\over\del\Phi^B(u',\tau')},\nn\\
&&\ve\left({\del\over\del\Phi^A}\right)=
\ve\left({\ldel\over\del\Phi^A}\right)=\ve_A+n,\nn\\&&
F{\stackrel{\leftarrow}{\del}\over
\del\Phi^A(u,\tau)}d^n\tau=(-1)^{\ve_A(\ve(F)+1)}d^n\tau
{\del\over\del\Phi^A(u,\tau)}F,
\e{7}
where the delta-function in the odd coordinates $\tau^a$ satisfies
 the
properties
\be
&&\int f(\tau')\del^n(\tau-\tau')d^n\tau'=f(\tau)=\int
d^n\tau'\del^n(\tau'-\tau)f(\tau').
\e{101}

 An explicit calculation of the antibracket of the master actions \r{1} yields
\be
&&\half(\Sigma,\Sigma)=\int d^{n}u
d^{n}\tau\left(D\cL_{n}(u,\tau)+\half(S,S)_n(u,\tau)\right)
\e{11}
where we have introduced the local $n$-bracket
\be
&(f, g)_n\equiv&
f{\ldif\over\dif\Phi^A}
{\stackrel{\rightarrow}{\dif}\over\dif\Phi^*_A}
g-(f\leftrightarrow g)(-1)^{(\ve(f)+1+n)(\ve(g)+1+n)},
\e{13}
where $f$ and $g$ are local functions of $\Phi^A(u,\tau)$ and
$\Phi^*_A(u,\tau)$.
For even $n$ this is an antibracket and for odd $n$ it is a Poisson
bracket. In other
words $\Phi^A$ and
$\Phi_A^*$ are canonical conjugate field variables on an antisymplectic
manifold for
even  $n$ and on a symplectic manifold for odd $n$. The condition  \r{4}
combined
with \r{11} yields the conditions
\be
&&\int d^nud^n\tau D\cL_n(u,\tau)=0,
\e{141}
which determines the allowed boundary conditions,
and
\be
&&(S,S)_n=0.
\e{15}
Furthermore, we have from \r{1} and \r{6}
\be
&&\Delta \Sigma =0
\e{16}
since the $\tau$-part yields a factor zero. (As usual we believe that the
bosonic part may be regularized appropriately.) $\Sigma$ satisfies
therefore also the
quantum master equation
\be
&&\half(\Sigma,\Sigma)=i\hbar\Delta \Sigma,
\e{17}
when \r{141} and \r{15} are satisfied,
which means that no quantum corrections of the measure in the path integral are
required for these models.

If one treats the master action $\Sigma$ in \r{1} as an ordinary action,
then the
equations of motion are
\be
&&D\Phi^A=(S,\Phi^A)_n,\quad D\Phi^*_A=(S,\Phi^*_A)_n,
\e{18}
the consistency of which again requires \r{15}. Notice that the equation of
motion
for $S$ then is $DS=0$. Notice also the relations 
\be
&&(\Sigma, \Phi^A)=(-1)^n\left(D\Phi^A-(S, \Phi^A)_n\right),\nn\\&&
(\Sigma, \Phi^*_A)=(-1)^n\left(D\Phi^*_A-(S, \Phi^*_A)_n\right).
\e{181}

The superfields above may be Taylor expanded in the odd $\tau$-coordinates
in such a
fashion that the coefficient fields are fields and antifields in the
ordinary sense
on the base manifold. Or in other words such that the antibracket \r{5} and the
$\Delta$-operator \r{6} have their conventional forms:
\be
&&(F, G)= \sum_r\int
F{\ldel\over\del\Phi^{rA}(u)} d^nu
{\stackrel{\rightarrow}{\del}\over\del\Phi^*_{rA}(u)}G-(F\leftrightarrow
G)(-1)^{(\ve(F)+1)(\ve(G)+1)},\nn\\
&&\Delta\equiv\sum_r\int d^nu
(-1)^{\ve_{rA}}{\del\over\del\Phi^{rA}(u)}{\del\over\del\Phi^*_{rA}(u)},\quad
\ve_{rA}\equiv\ve(\Phi^{rA}),
\e{20}
where $\Phi^{rA}(u)$ and $\Phi^*_{rA}(u)$ are the coefficient fields ($r$
denotes
antisymmetric $u$-indices). The antibracket in \r{20} must yield
\be
&&(\Phi^A(u,\tau),
\Phi^*_B(u',\tau'))=(-1)^{n\ve_A}\del^A_B\del^{n}(u-u')\del^{n}(\tau-\tau'),
\e{21}
which trivially follows from \r{5}. An explicit construction is given in appendix B.

So far we have given superfield actions in arbitrary dimensions which under
certain
conditions satisfy the master equations. In the following we will always require
$\Sigma$ to actually satisfy the master equation, \ie we require the boundary
conditions to be consistent with
\r{141} and the local equation \r{15} to be satisfied. In order to
determine  the
class of gauge field theories  which have such master
actions   we need also to prescribe ghost numbers to the field. We
generalize then the
prescription given in
\cite{Sigma} (such ghost numbers were also considered in \cite{CF,GD,Park,BBD}):
 We choose the odd coordinates
$\tau^a$ to have ghost number plus one, which implies that $D$ in \r{3}  has
ghost number plus one.  Since we require the master action $\Sigma$ to have
ghost
number zero and since the measure $d^n\tau$ has ghost number $-n$ the Lagrangian
density
$\cL_n$ in \r{2} must have ghost number $n$, \ie
\be
&&gh(\Sigma)=0,\quad gh(d^n\tau)=-n,\;\;\Rightarrow\;\;gh(\cL_n)=n.
\e{3311}
The form \r{2} of $\cL_n$  leads us to the following
general rule for the superfields and the local function
$S$:
\be
&&gh(\Phi^A)+gh(\Phi^*_A)=n-1, \quad gh(S)=n.
\e{34}
We assume that $S$ is given by a power expansion in the superfields $\Phi^A$ and
$\Phi^*_A$ in which case this ghost number prescription will restrict the
possible
terms in $S$. In such an analysis it is convenient to use the following
convention:
\be
&&gh(\Phi^*_A)\geq gh(\Phi^A),
\e{36}
which in itself does not impose any restriction.
Notice that $\Phi^A$ and
$\Phi^*_A$  can only have equal ghost number in odd dimensions. (More
precisely for
$n=2m+1$ we may have the ghost numbers
$gh(\Phi^A)=gh(\Phi^*_A)=m$.)
Since the local bracket \r{13} satisfies
\be
&&(\Phi^A(u,\tau),
\Phi^*_B(u,\tau))_n=\del^A_B,
\e{361}
it follows from the ghost number prescription \r{34} that the local bracket
$(\;,\;)_n$ in itself carries ghost number $1-n$. From \r{21} and the fact that
$\del^n(\tau-\tau')$ carries ghost number $n$, it follows that the
antibracket \r{5}
or equivalently \r{20} carries ghost number plus one which is the standard ghost
number prescription for the antibracket in the field-antifield formalism.

When we for a given set of superfields which satisfy \r{34} and   boundary
conditions consistent with
\r{141} also find an
$S$ satisfying  $gh(S)=n$ and  $(S,S)_n=0$, then we have a
consistent master action $\Sigma$ given by \r{1}. From such a  master
action we may
then extract the original model according to the following rules: First
remove the
integration measure. Then perform the following replacements:
\be
&&D\;\longrightarrow\;{\rm exterior}\;{\rm derivative}\;d\nn\\
&&\Phi^A, \;gh(\Phi^A)=k\geq0\;\longrightarrow\;k{\rm -form}\;{\rm
field}\;\phi^A,\;\ve(\phi^A)=\ve_A+k,\nn\\
&&\Phi^*_A, \;gh(\Phi^*_A)=(n-1-k)\geq0\;\longrightarrow\;(n-1-k){\rm -form}\;{\rm
field}\;\phi^*_A,\nn\\&&\hspace{6cm}\ve(\phi^*_A)=\ve_A+k,\nn\\
&&\Phi^A,
\;gh(\Phi^A)<0,\;\longrightarrow\;0,\quad {\rm or}\quad
\Phi^*_A,\;gh(\Phi^*_A)<0,\;\longrightarrow\;0,\nn\\
&&{\rm ordinary}\;{\rm multiplication}\;\longrightarrow\;{\rm wedge}\;{\rm
products},
\e{37}
where the form fields, $\phi^A$ and $\phi^*_A$, are fields on the $n$-dimensional $u$-space.
These rules
are easily extracted from the fact that the original fields, $\phi^A(u)$ and $\phi^*_A(u)$, are component fields
with ghost number zero  of the superfields, $\Phi^A(u,\tau)$ and $\Phi^*_A(u,\tau)$. (Signs depend on how the coefficient fields are defined
precisely.)

The superfield master actions determine also the gauge invariance of the
original
action. The appropriate gauge transformations may be obtained as follows: First
identify  the components of the superfields that are  antifields to the original
fields. These antifields have ghost number minus one and if $\Phi^A$
contains the
field then $\Phi^*_A$ contains the corresponding antifield or vice versa.
Identify then the terms in the master Lagrangian $\int d^n\tau\cL_n$ which
are linear
in these antifields. The terms in the coefficients of these antifields which
contain no antifields represent then the gauge transformations of the
field. The ghost
fields with ghost number plus one are then to be interpreted as gauge
parameters. 

\section{Master actions in terms of general superfields}
In \cite{Sigma} we generalized the two-dimensional treatment along the
lines of the
generalized antisymplectic formulation in \cite{BT}. Here we give the
corresponding
treatment in arbitrary dimensions. Let the superfields $Z^I(u,\tau)$,
$\ve(Z^I)\equiv\ve_I$, represent arbitrary coordinates on a  (anti)symplectic
manifold. The local brackets \r{13} are then defined by
\be
&&(f, g)_n\equiv
f(Z(u,\tau)){\ldif\over\dif Z^I}E^{IK}(Z(u,\tau))
{\stackrel{\rightarrow}{\dif}\over\dif Z^K}
g(Z(u,\tau)),
\e{38}
which implies
\be
&&(Z^I(u,\tau), Z^K(u,\tau))_n=
E^{IK}(Z(u,\tau)).
\e{39}
The functions $E^{IK}$ have the properties
\be
&&\ve(E^{IK})=\ve_I+\ve_K+1+n,\quad E^{KI}=-E^{IK}(-1)^{(\ve_I+1+n)(\ve_K+1+n)}.
\e{40}
These properties follow from the requirements
\be
&&\ve((f,g)_n)=\ve_f+\ve_g+1+n,\quad
(f,g)_n=-(g,f)_n(-1)^{(\ve_f+1+n)(\ve_g+1+n)}.
\e{41}
The Jacobi identities of the bracket \r{38}, \ie
\be
&&((f,g)_n,h)_n(-1)^{(\ve_f+1+n)(\ve_h+1+n)}+cycle(f,g,h)=0,
\e{411}
require in turn
\be
&&E^{IL}\dif_LE^{JK}(-1)^{(\ve_I+1+n)(\ve_K+1+n)}+cycle(I,J,K)=0.
\e{42}
We assume that $Z^I$ span the bracket \r{38} which implies that $E^{IK}$ is
invertible. The inverse, $E_{IK}$, is defined by
\be
&&E_{IL}E^{LK}=\del^K_I=E^{KL}E_{LI}.
\e{43}
It satisfies the properties
\be
&&\ve(E_{IK})=\ve_I+\ve_K+1+n,\quad
E_{KI}=-E_{IK}(-1)^{(\ve_I+n)(\ve_K+n)+n},\nn\\
&&\dif_IE_{JK}(Z)(-1)^{(\ve_I+n)\ve_K}+cycle(I,J,K)=0.
\e{44}
The last relation follows  from \r{43} and the Jacobi identities \r{42}.

We define the generalized antibracket by
\be
&&(F, G)\equiv
F\int  {\ldel\over\del Z^I(u,\tau)}d^{n}u
d^{n}\tau
(Z^I(u,\tau), Z^K(u',\tau'))d^{n}u'
d^{n}\tau'{\stackrel{\rightarrow}{\del}\over\del Z^K(u',\tau')}G,\nn\\
\e{45}
where
\be
&&(Z^I(u,\tau), Z^K(u',\tau'))=
E^{IK}(Z(u,\tau))(-1)^{n\ve_K}\del^n(u-u')\del^n(\tau-\tau')=\nn\\&&=\del^n(u-u')\del^n(\tau-\tau')
(-1)^{n\ve_I}E^{IK}(Z(u,\tau)).
\e{46}
The Jacobi identities are here
\be
&&(Z^I(u,\tau), (Z^J(u',\tau'),
Z^K(u'',\tau'')))(-1)^{(\ve_I+1)(\ve_K+1)}\nn\\&&+cycle(I,u,\tau;J,u',\tau';
K,u'',\tau'')=0,
\e{462}
which are satisfied due to \r{42}. In fact, the properties of $E^{IK}$
imply that
the antibracket \r{45}, \r{46}
satisfies all required properties. Inserting \r{46} into \r{45} yields the
expression
\be
&&(F, G)\equiv
F\int  {\ldel\over\del Z^I(u,\tau)}d^{n}u
d^{n}\tau
E^{IK}(-1)^{n\ve_K}{\stackrel{\rightarrow}{\del}\over\del Z^K(u,\tau)}G.
\e{463}
This antibracket follows also from the relation \r{102} and the
$\Delta$-operator
\be
&&\Delta\equiv\half\int d^nu d^n\tau
(-1)^{\ve_I}\rho^{-1}{{\del}\over\del Z^I(u,\tau)}\rho
E^{IK}(Z(u,\tau))(-1)^{n\ve_K}{{\del}\over\del Z^K(u,\tau)},\nn\\
\e{461}
where $\rho(Z(u,\tau))$ is a measure density on the field-antifield space. This
$\Delta$-operator differentiates the antibracket \r{463} according to the rules
\r{601}.

A master action $\Sigma$ in terms of the superfields $Z^I(u,\tau)$ may \eg
have the
form \r{1} with a Lagrangian density given by
\be
&\cL_n(u,\tau)&=V_I(Z(u,\tau))D Z^I(u,\tau)(-1)^{\ve_I+n}-
S(Z(u,\tau)),
\e{47}
where the
 (anti)symplectic potential $V_I(Z(u,\tau))$ has the Grassmann parity
$\ve_I+n+1$.
Here we obtain the left functional derivative of $\Sigma$ in the form:
\be
&&{\del\Sigma\over\del Z^I}=(-1)^{n\ve_I}\left(E_{IK}DZ^K(-1)^{\ve_K+n}-\dif_IS\right),
\e{471}
where 
$E_{IK}$ is defined by
\be
&&E_{IK}(Z)=\dif_IV_K(Z)-\dif_KV_I(Z)(-1)^{(\ve_I+n)(\ve_K+n)+n}.
\e{49}
The equations of motion are therefore
\be
&&DZ^I=(S, Z^I)_n.
\e{48}
Consistency requires then again $(S,S)_n=0$.
A still more general form for the master action $\Sigma$ is obtained from the
Lagrangian density
\be
&\cL_n(u,\tau)&=Z^K(u,\tau)\bar{E}_{KI}(Z(u,\tau))DZ^I(u,\tau)(-1)^{\ve_I+n}
-S(Z(u,\tau)),
\e{491}
where
\be
&&\bar{E}_{KI}(Z)\equiv (Z^J\dif_J+2)^{-1}E_{KI}(Z)=\int_0^1d\al\al
E_{KI}(\al Z).
\e{492}
(Such an expression  was first given in the symplectic case in \cite{BF}.) Notice also the general form of the expression \r{181}:
\be
&&(\Sigma, Z^I)=(-1)^n\left(DZ^I-(S,Z^I)_n\right).
\e{4921}

When calculating $(\Sigma,\Sigma)$ in terms of the antibracket \r{45} or
\r{463},
where
$\Sigma$ is expressed in terms of \r{47} or \r{491}, we again find the relation
\r{11}. This means that the master equation
$(\Sigma,\Sigma)=0$  also here requires  the conditions \r{141}, \ie $\int
d^nud^n\tau D\cL_n=0$, which determines the allowed boundary conditions, and
$(S,S)_n=0$ in terms of the bracket \r{38} which together with the ghost
number prescription determine the allowed form of
$S$.

The ghost number prescriptions are here
\be
&&gh(S)=n, \quad gh(V_I)+gh(Z^I)=n-1,\; {\rm each}\; I,
\e{493}
for the Lagrangian \r{47}, and
\be
&&gh(S)=n, \quad gh(Z^K)+gh(Z^I)+gh(\bar{E}_{IK})=n-1,\; {\rm each}\; I\;
{\rm and}\;
K,
\e{494}
for the Lagrangian \r{491}. The last relation together with \r{492}, or
\r{493} and
\r{49} imply
\be
&&gh(E_{IK}(Z))=n-1-gh(Z^I)-gh(Z^K),\nn\\&&
gh(E^{IK}(Z))=1-n+gh(Z^I)+gh(Z^K),
\e{495}
where the last relation follows from \r{43}. The last equality implies then
that the
local bracket $(\;,\;)_n$ carries the extra ghost number $1-n$ (see \r{39})
exactly
what we had in the previous section. Only for
$n=1$ do we have the  standard Poisson bracket.

\section{Further generalizations}
1) As in \cite{Sigma} we may choose arbitrary coordinates $u^a$ on the
surface. For
the measure
\be
&&d^nu\left(\det h^b_a(u)\right)^{-1}
\e{5000}
we have the nilpotent $D$-operator
\be
&&D=\tau^aT_a+\half\tau^b\tau^aU^c_{ab}(u){\dif\over\dif\tau^c},\quad
\e{5100}
where
\be
&&T_a\equiv
h^b_a(u){\dif\over\dif u^b}, \quad
U^c_{ab}(u)\equiv-h^f_a(u)h^d_b(u){\dif\over\dif
u^f}\left(h^{-1}\right)^c_d(u)-(a\leftrightarrow b).
\e{52}
Notice that
\be
&&D^2=0\;\Rightarrow\;[T_a, T_b]=U^c_{ab}(u)T_c.
\e{53}\\ \\
2) If the original field theory is a superfield theory on a space with
coordinates
$u^a$ having Grassmann parities $\ve_a$, then the $\tau^a$-coordinates have
Grassmann
parities
$\ve_a+1$. In this case we must assume that the superfields
$\Phi^A(u,\tau)$ still
are possible to expand as a power series in $\tau^a$ which then is an infinite
expansion for the bosonic $\tau^a$-coordinates. (The fact that this implies an
infinite number of component fields with arbitrarily low ghost numbers
suggests that
 infinite reducibility is a generic feature in the superfield case.) If the
original
base manifold has dimension
$(n,m)$ then the supermanifold has dimension  $(n+m,n+m)$ and we get a
subdivision
into the two cases
$n+m$ odd or even. This means that the dimension of the base space still
determines
the odd and even cases.\\
\\ 3) The local functions $S$ may also have explicit $\tau$-dependence, since
the master actions above  still satisfy the master equations provided $S$
satisfies
\r{15}. However, such master actions are not of the same geometric nature as the
previous ones since they do not lead to topological field theories due to the
terms with explicit $\tau$-dependence.  (This generalization was
first considered in the one-dimensional treatments in
\cite{GD,BBD}.)\\ \\
4) A natural  extension of the formalism in section 2 is to let $E^{IK}(Z)$ be singular. It may \eg be a Dirac bracket in which case we have 
\be
&&E^{IK}_{({\cD})}\equiv(Z^I, Z^K)_{n(\cD)}\equiv (Z^I, Z^K)_n-(Z^I, \Theta^{\mu}(Z))_nC_{\mu\nu}(\Theta^{\nu}(Z), Z^K)_n,
\e{50}
where $\Theta^{\mu}(Z)$ are constraint variables such that $C^{\mu\nu}\equiv(\Theta^{\mu}, \Theta^{\nu})_n$ is invertible with the inverse $C_{\mu\nu}$. The basic Dirac antibracket is then according to \r{46} given by
\be
&&(Z^I(u,\tau), Z^K(u',\tau'))_{(\cD)}=E^{IK}_{(\cD)}(Z(u,\tau))(-1)^{n\ve_K}\del^n(u-u')\del^n(\tau-\tau').
\e{51}
The local action $S$ should be replaced by
\be
&&S(Z(u,\tau))\;\longrightarrow\;S(Z(u,\tau))+\Pi_{\mu}(u,\tau)\Theta^{\mu}(Z(u,\tau)),
\e{5111}
where $\Pi_{\mu}$ is a Lagrange multiplier superfield.

\section{Generalized Poisson sigma models in any dimension.}
Consider dimension $n$ and consider first superfield pairs $X^i$ and $X^*_i$ with the
properties
\be
&&\ve(X^*_i)=\ve_i+1+n,\quad\ve_i\equiv\ve(X^i),\nn\\
&&gh(X^*_i)=n-1-gh(X^i),
\e{54}
which are required by our general rules in section 1. Consider then the
local function
\be
&&S=\half X^*_jX^*_i\om^{ij}(X)(-1)^{\ve_j+n}.
\e{55}
From the general prescription
\be
&&\ve(S)=n,\quad gh(S)=n,
\e{56}
it follows that the functions $\om^{ij}(X)$ must satisfy the properties
\be
&&\ve(\om^{ij})=n+\ve_i+\ve_j,\nn\\
&&gh(\om^{ij})=2-n+gh(X^i)+gh(X^j).
\e{57}
The expression \r{55} implies furthermore that
\be
&&\om^{ij}=-\om^{ji}(-1)^{(\ve_i+n)(\ve_j+n)}.
\e{58}
The master equation $(S,S)_n=0$ requires then
\be
&&\om^{il}\dif_l\om^{jk}(-1)^{(\ve_i+n)(\ve_k+n)}+cycle(i,j,k)=0.
\e{59}
From these results it follows that $\om^{ij}(X)$ has exactly the same
properties as
the local bracket
\r{38} for $n-1$. Therefore,  we may make the identification
\be
&&\om^{ij}(X)=(X^i, X^j)_{n-1},
\e{591}
where $(\;,\;)_0$ may be identified with the conventional antibracket in the
field-antifield formalism. For even $n$ $\om^{ij}$ is a Poisson bracket and
for odd
$n$ $\om^{ij}$ is an antibracket. But only for $n=1,2$ do these brackets
have the
conventional ghost number prescriptions. ($(\;,\;)_1$ is the conventional
Poisson
bracket.) The boundary conditions must be consistent with \r{141}.

When $\om^{ij}(X)$ satisfies the above properties and when the boundary
conditions are
consistent with \r{141} then the action
\be
&&\Sigma[X,X^*]=\int d^nu d^n\tau \left( X^*_iDX^i(-1)^{\ve_i+n}-S\right)
\e{592}
satisfies the master equation $(\Sigma, \Sigma)=0$.
 However, for this case the original models may only be written down after
we have
given a more explicit form for
$\om^{ij}(X)$. The only exception is for $n=2$ and $gh(X^i)=0$ in which case,
 according to the rules \r{37}, we
directly obtain
 \be
&&A=\int\left(x^*_i\wedge
dx^i-\half x^*_j\wedge x^*_i\om^{ij}(x)\right),
\e{65}
 where $x^*_i$ and
$x^i$ are one-form and zero-form fields respectively. (The Grassmann parities
are $\ve(x^*_i)=\ve(x^i)\equiv\ve_i$.) This is just the
well-known Poisson sigma model
\cite{Ikeda,SS} for which Cattaneo and Felder also gave the superfield master
action given here for $\ve(x^i)=0$ \cite{CF}. One may easily check that the
boundary
conditions given in \cite{CF} are consistent with \r{141}.

In
\cite{Sigma}  generalized Poisson sigma models for $n=2$ were constructed
by means of
the algorithm presented here. This construction may also be generalized to
arbitrary
$n$. We consider then the following expression for $S$: 
\be
&&S=\half X^*_jX^*_i\om^{ij}(X)(-1)^{\ve_j+n}+\Lambda^*_{\al}\theta^{\al}(X),
\e{651}
where we have introduced new superfield pairs
$\Lambda^*_{\al}$,
$\Lambda^{\al}$ with the properties
\be
&&\ve(\Lambda^*_{\al})=\ve_{\al}+n,\quad \ve(\Lambda^{\al})=\ve_{\al}+1,\quad \ve_{\al}\equiv \ve(\theta^{\al}),\nn\\&&
gh(\Lambda^{\al})=gh(\theta^{\al})-1, \quad gh(\Lambda^*_{\al})=n-gh(\theta^{\al}).
\e{6511}
 The local master
equation $(S,S)_n=0$ requires now apart from \r{59} also
\be
&&\theta^{\al}(X)\ldif_j\om^{ji}(X)=0,
\e{652}
which means that $\om^{ji}(X)$  may be interpreted as a Dirac bracket of
the $n-1$
type, \ie a bracket $(\cdot,\cdot)_{n-1}$ satisfying
$(\cdot,\theta^{\al}(X))_{n-1}=0$.

The expression \r{651} is
not the general expression of $S$ in terms of these fields. We may also add
the terms
\be
&&(-1)^{\ve_j+n+(\ve_i+n)(\ve_k+n)}{1\over
6}X^*_iX^*_jX^*_k\om^{kji}_{\al}(X)\La^{\al}+
(-1)^{\ve_i+n}X^*_i\La^*_{\al}\om^{\al
i}_{\beta}(X)\La^{\beta}+\ldots,\nn\\
\e{653}
where 
\be
&&\ve(\om^{kji}_{\al}(X))=\ve_i+\ve_j+\ve_k+\ve_{\al},\quad \ve(\om^{\al
i}_{\beta}(X))=\ve_i+\ve_{\al}+\ve_{\beta}+n,\nn\\
&&gh(\om^{kji}_{\al}(X))=gh(X^i)+gh(X^j)+gh(X^k)-gh(\theta^{\al})+4-2n,\nn\\&& gh(\om^{\al
i}_{\beta}(X))=gh(X^i)+gh(\theta^{\al})-gh(\theta^{\beta})+2-n,
\e{661}
and where
$\om^{kji}_{\al}$ is totally antisymmetric in $i$, $j$, $k$ with Grassmann
parity\\
$\ve_{ijk}=(\ve_i+n)(\ve_j+n)+(\ve_j+n)(\ve_k+n)+(\ve_k+n)(\ve_i+n)$.
In this case the master equations yield a weak form of the Jacobi
identities \r{59}
and the degeneracy condition \r{652}, which means that $\om^{ij}(X)$ now is
a weak
Dirac bracket of the
$n-1$ type. For all these forms of $S$ the master action is given by
\be
&&\Sigma[X,X^*,\Lambda,\Lambda^*]=\int d^nud^n\tau\left(X^*_iDX^i(-1)^{\ve_i+n}+\Lambda^*_{\al}D\Lambda^{\al}(-1)^{\ve_{\al}+n+1}-S\right).\nn\\
\e{6611}

We may also add further superfield pairs $\Xi^{\al_k}$,
$\Xi^*_{\al_k}$,
$k=1,\ldots,L$, which enter $S$ to leading order in the form $\sum_{k=0}^{L-1}\Xi^*_{\al_{k+1}}Z^{\al_{k+1}}_{\al_k}(X)\Xi^{\al_k}$ ($k\geq0$, $\al_0\equiv\al$, $\Xi^{\al_0}\equiv\Lambda^{\al}$) where
\be
&& \ve(Z^{\al_{k+1}}_{\al_k})=\ve_{\al_{k+1}}+\ve_{\al_k},\quad\ve(\Xi^{\al_k})\equiv\ve_{\al_k}+k+1, \nn\\
&&
 gh(Z^{\al_{k+1}}_{\al_k})=gh(\Xi^{\al_{k+1}})-gh(\Xi^{\al_k})+1.
\e{662}
The master action is here
\be
&&\Sigma[X,X^*,\Xi,\Xi^*]=\nn\\&&=\int d^nud^n\tau\left(X^*_iDX^i(-1)^{\ve_i+n}+\sum_{k=0}^{L}\Xi^*_{\al_{k}}D\Xi^{\al_k}(-1)^{\ve_{\al_k}+k+1+n}-S\right),
\e{6621}
and the master equations tell us that $\theta^{\al}$ is reducible to stage $L$, which means $Z^{\al_1}_{\al}\theta^{\al}=0, Z^{\al_3}_{\al_2}Z^{\al_2}_{\al_1}\approx 0, \ldots, Z^{\al_L}_{\al_{L-1}}Z^{\al_{L-1}}_{\al_{L-2}}\approx 0$.

The original model
and its properties of all these modified forms of $S$ can only be written
down when we
have given a more explicit form for the ghost dependent functions. However,
for $n=2$
and
$gh(X^i)=0$
 we
 obtain
 according to the rules
\r{37} for all the above cases the original action
\be
&&A=\int\left(x^*_i\wedge
dx^i-\half x^*_j\wedge x^*_i\om^{ij}(x)-
\la_{\al}\theta^{\al}(x)\right),
\e{66}
where $\la_{\al}$ is a Lagrange multiplier $2$-form field.

Still more generalized forms of Poisson sigma models might be possible to
derive
if we allow $S$ to have explicit $\tau$-dependence which is possible
according to the
generalization 3 in section 3.
\\
\\

\noindent
{\bf Acknowledgements}:

I.A.B. would like to thank Lars Brink for
his very warm hospitality at the
Department of Theoretical Physics, Chalmers
and G\"oteborg University.
 The work of I.A.B. is supported by the grants 99-01-00980 and
99-02-17916 from Russian foundation for basic researches and by the
President grant
00-15-96566 for supporting leading scientific schools. This work is partially
supported by the grant INTAS 00-00262.

\begin{appendix}
\section{Further applications}
\subsection{Models in $n=1$}
Consider first superfield pairs $X^i$ and $X^*_i$ with
ghost number zero and arbitrary Grassmann parities $\ve(X^*_i)=\ve(X^i)\equiv \ve_i$.
If these are
the only superfields  then
$S=0$,  since
$S$ must have ghost number one. The original model is then trivial and is
of the form
$A=\int x^*_idx^i$. If we also have the superfield pairs
$\La^*_{\al}$ and $\La^{\al}$ with ghost numbers one and minus one
respectively, then
$S$ has the general form ($\ve(\La^*_{\al})=\ve(\Lambda^{\al})=\ve_{\al}+1$,
$\ve_{\al}\equiv\ve(\theta^{\al})$)
\be
&&S=\La^*_{\al}\theta^{\al}(X,
X^*)+\half\Lambda^*_{\al}\Lambda^*_{\beta}U^{\beta\al}_{\ga}(X,X^*)
\Lambda^{\ga}(-1)^{\ve_{\al}}+\ldots,
\e{60}
where the dotted terms are determined by the condition $(S,S)_1=0$. Notice
that $S$
is odd and that $(\;,\;)_1$ is a conventional  Poisson bracket. In fact,
\be
&&(S,S)_1=0\quad\Rightarrow\quad(\theta^{\al},\theta^{\beta})_1
=U^{\al\beta}_{\ga}\theta^{\ga}.
\e{61}
Thus, the master action 
\be
&&\Sigma[X,X^*,\Lambda,\Lambda^*]=\int dud\tau\left(X^*_iDX^i(-1)^{\ve_i+1}+\Lambda^*_{\al}D\Lambda^{\al}(-1)^{\ve_{\al}}-S\right),
\e{611}
where $S$ is given in \r{60} satisfies the master
equation provided
the constraint variables $\theta^{\al}$ are in involutions. We may also add
further
pairs
$\Xi^*_{\al_k}$, $\Xi^{\al_k}$ with ghost number $k>1$ and $-k$. The
resulting
$S$ satisfying $(S,S)_1=0$ will then imply that the constraint variables
$\theta^{\al}$ are reducible (linearly dependent) up to a certain stage.
Thus, $S$
may attain the general form of a BFV-BRST charge for a constraint theory
where the
constraints are in arbitrary involutions on the phase space spanned by the
canonical
coordinates
$X^i$ and
$X^*_i$.
$\La^*_{\al}$ are ghosts and $\Xi^*_{\al_k}$ ghost for ghosts.
The corresponding BRST
charge for the original model is the $\tau=0$ component of the $S(u,\tau)$
in \r{60}.
The boundary condition
\r{141} requires the conservation of this charge, \ie $S(u_2,0)-S(u_1,0)=0$
where
$u_1$ and
$u_2$ are the limits of the $u$-integration in $\Sigma$.
 The original action is obtained from the corresponding
master action using the rules \r{37}. We find for all the above cases the
original
action
\be
&&A=\int
\left(x^*_idx^i-\la_{\al}\theta^{\al}(x, x^*)\right),
\e{62}
where $\la_{\al}$ is a Lagrange multiplier one-form field. This action  is
defined on the phase space where $x^*_i$ and $x^i$ are canonical conjugate
variables ($x^*_i$ are conjugate momenta to $x^i$).
 It is gauge invariant
under the gauge transformations
\be
&&\del x^i=(x^i, \theta^{\al})_1\beta_{\al},\quad \del
x^*_i=(x^*_i, \theta^{\al})_1\beta_{\al},\nn\\&&
\del\la_{\al}=d\beta_{\al}(-1)^{\ve_{\al}}-
\beta_{\mu}\la_{\nu}U^{\nu\mu}_{\al}(-1)^{\ve_{\mu}},
\e{63}
where $\beta^{\al}$ are the gauge parameters.
 The theory \r{62} is an arbitrary constraint theory with zero
Hamiltonian, which implies that the theory is reparametrization invariant.
In fact,
any one-dimensional Hamiltonian model may be cast into this form, since any such
theory may be cast into a reparametrization invariant form. However, a nonzero
Hamiltonian
$H(X, X^*)$ is possible to introduce if we  allow $S$ to have an explicit
$\tau$-dependence  (generalization 3 in section 3, see also
\cite{GD}). Condition \r{15}  requires then  $H(X, X^*)$ to have zero
local Poisson bracket with the $S$ in \r{60}.
We may also choose fields which are  arbitrary symplectic coordinates if
we make use of the general master actions following from \r{47} and \r{491}.
These results  agree with  those obtained in
\cite{GD}.

\subsection{The Chern-Simon model and generalizations}
In $n=3$ the natural choice of  superfield pairs are  $X^i$ and $X^*_j$ both
with  ghost number one. Since $S$ must have ghost number three, $S$ must be
trilinear
in
$X^i$ and $X^*_j$. This case is simpler to analyze if we let the superfield
$Z^I$ represent both $X^i$ and $X^*_j$.
$Z^I$ are then Darboux coordinates on a symplectic manifold.
The local Poisson bracket is then
\be
&&(Z^I, Z^K)_3=E^{IK},
\e{70}
where $E^{IK}$ is constant. (Notice that the bracket $(\;,\;)_3$ carries an
extra ghost number $-2$ which implies $gh(E^{IK})=0$ for $gh(Z^I)=1$.) The
general
form of $S$ is in this case
\be
&&S={1\over 6} C_{IJK}Z^IZ^JZ^K.
\e{71}
If we let the fields $Z^I$ be odd fields corresponding to original fields
which are even one-form fields, then
 $C_{IJK}$ in \r{71} are even real constants. Notice also that $E^{IK}$ in
\r{70} is
symmetric for odd fields. The master action with the Lagrangian density \r{47}
becomes in this case
\be
&&\Sigma[Z]=\half\int d^3ud^3\tau\left(E_{IK}Z^KDZ^I-{1\over
3}C_{IJK}Z^IZ^JZ^K\right).
\e{72}
The master equation $(\Sigma, \Sigma)=0$ allows us then to interpret
$C_{IJK}$ as
structure coefficients of a Lie group, since $C_{IJK}$ is totally antisymmetric
from \r{71} and since
$(S,S)_3=0$ requires them to satisfy the Jacobi identities. $E^{IK}$ acts
as a group
metric. According to the rules
\r{37} the original model is
\be
&&A=\half\int E_{IK}z^I\wedge\left(dz^K-{1\over
3}E^{KL}C_{LMN}z^M\wedge z^N\right),
\e{73}
which is a non-abelian Chern-Simon model. $z^I$ are  even, one-form
fields. (The
Chern-Simon model was also treated in \cite{AKSZ}.)

 We may also introduce superfield pairs
 with ghost numbers two and zero
respectively, and  pairs
with ghost numbers three and minus one. These
superfields allow for many more terms in $S$ which are consistent with the ghost
number prescription. It seems likely that the local master equation \r{15}
then will
allow for new nontrivial solutions, in which case one will find generalized
Chern-Simon models. All fields must  satisfy boundary conditions which are
consistent with
\r{141}.

In  $n=4$ we may obtain something similar to the Chern-Simon model if we choose
 superfield pairs  $X^i$ and $X^*_i$ with
ghost number one and two respectively. It is then natural to let $X^i$ be
odd and
$X^*_i$ even since the
corresponding original one- and two-form fields, $x^i$ and $x^*_i$,
then are
even according to the rule \r{37}.
The general form of
$S$ is then
\be
&S=&\half C^{ij}X^*_iX^*_j+\half
C^{i}_{\;\;jk}X^*_iX^jX^k+\nn\\&&+{1\over 24}
C_{ijkl}X^iX^jX^kX^l,
\e{80}
where the coefficients are even real constants. In this case the condition
$(S,S)_4=0$ yields the following three conditions:
\be
&&C^{i}_{\;\;jk}C^{kl}X^*_iX^*_lX^j=0,
\e{85}
\be
&&\left(C^{i}_{\;\;jm}C^{m}_{\;\;kl}+{1\over 3}C_{jklm}C^{mi}\right)
X^*_iX^jX^kX^l=0,
\e{83}
\be
&&C_{ijkn}C^{n}_{\;\;lm}X^iX^jX^kX^lX^m=0.
\e{81}
We notice the following special solutions:\\
i) If $C^{ij}=C_{ijkl}=0$ then $C^{i}_{\;\;jk}$ may be interpreted as structure
coefficients
of a Lie group since $(S,S)_4=0$ requires the Jacobi identities due to
\r{83}.\\ \\
ii) If only $C_{ijkl}=0$ then the same interpretation of $C^{i}_{\;\;jk}$ is
possible.
However, in this case $(S,S)_4=0$ requires not only the Jacobi identities for
$C^{i}_{\;\;jk}$ but also the condition
\be
&&C^{i}_{\;\;jl}C^{lk}+C^{k}_{\;\;jl}C^{li}=0,
\e{811}
from \r{85}.
If $C^{ij}$ is an invertible matrix, then $C_{ij}$ exists satisfying
$C^{ij}C_{jk}=\del^i_k$ and \r{811} may be rewritten as
\be
&&C_{ijk}+C_{kji}=0,\quad C_{ijk}\equiv C_{il}C^{l}_{\;\;jk}.
\e{821}
Thus, the symmetric matrix $C^{ij}$ may be interpreted as a group metric and
$C_{ijk}$ is totally antisymmetric as required by a semi-simple Lie
group. This
corresponds to what we had in the Chern-Simon model in three dimensions.
The original
model is according to the rule \r{37}:
\be
&&A=\int\left(x^*_i\wedge dx^i-\half
C^{ij}x^*_i\wedge x^*_j-\half
C^{i}_{jk}x^*_i\wedge x^j\wedge x^k\right),
\e{822}
where $x^i$ and $x^*_i$ are even one- and two-form fields
respectively. (They
are the ghost number zero coefficients of $X^i$ and $X^*_i$.) This action
is gauge
invariant under the transformations
\be
&&\del x^i=d\beta^i+C^{ij}\ga_j+C^{i}_{\;jk}\beta^j x^k,\nn\\
&&\del x^*_i=d\ga_i+C^{k}_{\;ij} x^*_k\beta^j+C^{k}_{\;ij}
\ga_k\wedge x^j,
\e{823}
where the gauge parameters $\beta^i$ and $\ga_i$ are zero- and one-forms
respectively.

Also here we may  introduce further superfield pairs with ghost numbers three and zero respectively and
pairs
with ghost numbers four and minus one
etc which
will allow for new terms
in
$S$. The local master equation \r{15} should then allow for new nontrivial
solutions
in which case one  obtains generalized models. Again all fields must satisfy
boundary conditions which are consistent with
\r{141}.

From the kinetic terms in the superfield Lagrangians \r{2} it is clear that
we get a
BF-theory in any dimension $n$, and  in any dimension we may also have a
Lagrangian
multiplier which is an $n$-form field in the original actions. However, in
higher and
higher dimensions there are more and more different $k$-form fields
allowed. The most
general structure will therefore be more and more  complex in higher and
higher dimensions.
  
\section{Superfields in terms of component fields}
A decomposition of the superfields in component fields satisfying \r{20} may be obtained by means of the following recursive formula. We start with zero components $\Phi^A_0$ and $\Phi^*_{0B}$ satisfying 
\be
&&(\Phi^A_0(u), \Phi^*_{0B}(u'))=\del^A_B\del^{n}(u-u').
\e{900}
We write then the superfields as follows
\be
&&\Phi^A(u,\tau^1,\ldots,\tau^{n-1}, \tau^n)=\Phi^A_0(u,\tau^1,\ldots,\tau^{n-1})+\tau^n\Phi^A_1(u,\tau^1,\ldots,\tau^{n-1}),
\e{901}
\be
&&\Phi^*_A(u,\tau^1,\ldots,\tau^{n-1}, \tau^n)=\nn\\&&=\left((-1)^{n-1}\Phi^*_{1A}(u,\tau^1,\ldots,\tau^{n-1})-\Phi^*_{0A}(u,\tau^1,\ldots,\tau^{n-1})\tau^n\right)(-1)^{\ve_A},
\e{902}
where
\be
&&(\Phi^A(u,\tau^1,\ldots,\tau^{n-1}), \Phi^*_B(u',\tau^{\prime 1},\ldots,\tau^{\prime n-1}))=\nn\\&&=\del^{n}(u-u')\del^{n-1}(\tau-\tau')\del^A_B(-1)^{(n-1)\ve_A}.
\e{903}
Then by construction the $n$-parametric superfields \r{901} and \r{902} satisfy the relations \r{903} with the formal replacement $n\rightarrow n+1$. One may easily chech that the following expressions satisfy these recursion formulas:
\be
&\Phi^A(u,\tau)=&\Phi^{A}_0(u)+\tau^a\Phi^{A}_a(u)+
\half\tau^{a_1}\tau^{a_2}\Phi^{A}_{a_1a_2}(u)+\cdots\nn\\&&\cdots+{1\over
k!}\tau^{a_1}\cdots\tau^{a_k}\Phi^A_{a_1\cdots a_k}(u)+\cdots +{1\over
n!}\tau^{a_{1}}\cdots\tau^{a_n}\Phi^A_{a_1a_2\cdots a_{n}}(u),\nn\\
&\Phi^*_A(u,\tau)=&\biggl({1\over n!}\Phi^{*a_1a_2\cdots a_{n}}_{A}(u)-
{1\over (n-1)!}\Phi^{*a_1a_2\cdots
a_{n-1}}_{A}(u)\tau^{a_{n}}+\nn\\&&+\half{1\over(n-2)!}\Phi^{*a_1a_2\cdots
a_{n-2}}_{A}(u)\tau^{a_{n-1}}\tau^{a_{n}}+\cdots\nn\\&&\cdots+(-1)^k{1\over
k!(n-k)!}\Phi^{*a_1\cdots
a_{n-k}}_A(u)\tau^{a_{n-k+1}}\cdots\tau^{a_n}+\cdots\nn\\&&\cdots
+(-1)^n{1\over
n!}\Phi^{*0}_A(u)\tau^{a_{1}}\cdots\tau^{a_n}\biggr)(-1)^{n\ve_A}\ve_{a_1a_2\cdots
a_{n}},\nn\\
\e{904}
where $\ve_{a_1a_2\cdots
a_{n}}$ is totally antisymmetric such that $\ve_{12\cdots
{n}}=1$. We also define\\$\ve_{1}\equiv 1$, $\ve_0\equiv 1$. Notice that $\Phi^A_{a_1\cdots a_k}$ and $\Phi_A^{*a_1\cdots a_k}$ are totally antisymmetric, and that we have
\be
&&(\Phi^A_{a_1\cdots a_k}(u), \Phi_B^{*a_1\cdots a_k}(u'))=\del^A_B\del^{n}(u-u').
\e{905}

\end{appendix}

\end{document}